\documentclass{aa501}
\usepackage{psfig}
\usepackage{lscape}
\usepackage{amsmath}
\usepackage{amssymb}
\usepackage[figuresleft]{rotating}

\newcommand{\MC}{\multicolumn}
\newcommand{\kms}{km~s$^\mathrm{-1}$}
\newcommand{\sunn}{$_{\odot}$}
\newcounter{qub}
\begin{document}

\title{
The metallicities of UM~151, UM~408 and A~1228+12 revisited
}

\author{S.A. Pustilnik\inst{1,4} \and
A.Y. Kniazev\inst{1,2,4} \and
J. Masegosa\inst{3}\thanks{Visiting Astronomer, German-Spanish
Astronomical Center, Calar Alto, operated by the Max-Planck-Institut
f\"{u}r Astronomie jointly with the Spanish National Commission for
Astronomy.} \and
I. M\'arquez\inst{3}$^*$\and
A.G. Pramsky\inst{1,4}  \and
A.V. Ugryumov\inst{1,4}
}

\offprints{S. Pustilnik, \email{sap@sao.ru}}

\institute{
Special Astrophysical Observatory RAS, Nizhnij Arkhyz,
Karachai-Circassia,  369167 Russia
\and Max Planck Institut f\"{u}r Astronomie, K\"{o}nigstuhl 17, D-69117,
Heidelberg, Germany
\and Instituto de Astrof\'{\i}sica de Andaluc\'{\i}a, Granada, Spain
\and Isaac Newton Institute of Chile, SAO Branch
}

 \date{Received December 5, 2001 / Accepted April 16, 2002}

 \abstract{
We present the results of new spectrophotometry and heavy element
abundance determinations for 3 dwarf galaxies  UM~151, UM~408 and
A~1228+12 (RMB~132). These galaxies have been claimed in the literature
to have very low metallicities, corresponding to $log(O/H)+12 \le$
7.65, that are in the metallicity  range of some candidate local young
galaxies. We present higher S/N data for these three galaxies.
UM~151 and UM~408 have significantly larger metallicities:  $log(O/H)+12 =$
8.5 and 7.93, respectively. For A~1228+12 our new $log(O/H)+12=$7.73 is close
to that recalculated from earlier data (7.68). Thus, the rederived
metallicities allow us to remove these objects from the list of
galaxies with $Z \lesssim 1/20~Z$\sunn.
   }

\authorrunning{S.A. Pustilnik et al.}

\titlerunning{The metallicities of UM~151, UM~408 and A~1228+12 revisited}

   \maketitle

\section{Introduction}

Extremely metal-deficient gas-rich galaxies (commonly taken as
objects with an ionized gas metallicity $Z \lesssim 1/20~Z_{\odot}$)
have received much attention recently. The discovery of an unusually
low metallicity of the ionized gas in the blue compact galaxy (BCG) I~Zw~18
 (Mkn~116) (Searle \& Sargent~\cite{SS72}) led to the claim that such
objects can be considered as the best candidates for truly
local young galaxies. However, the search for similar galaxies conducted by
several groups (e.g., Sargent \& Kunth~\cite{SK86}; Terlevich et al.
\cite{Terlevich91}) has failed to find a large
population with such extremely low metallicities.
In their review, Kunth \& \"Ostlin (\cite{KO}) summarize and discuss
many problems related to very metal-poor galaxies and they present a
compilation of all the known gas-rich galaxies with $Z \lesssim
1/20~Z_{\odot}$. From the analysis of the ratios of $C/O$ and $N/O$
abundances Izotov \& Thuan (\cite{IT99}) suggested that the lowest
metallicity blue compact galaxies should experience their first
episode of star formation (SF) and therefore the hypothesis of youth for
these galaxies needs to be carefully studied. In any case the properties
of such galaxies best approximate those of young low-mass galaxies formed at
the epoch of galaxy formation.

Therefore, very metal-poor galaxies in the Local Universe attract much
attention, and multiwavelength studies of such objects have been performed 
to better identify their nature.  However not all such galaxies found in the
early studies as very metal-deficient (Kunth \& \"Ostlin ~\cite{KO})
have spectroscopic data of sufficient quality for an accurate
determination of the metallicity.  Since detailed studies of individual
galaxies require significant effort and observational time, thus, one should
be confident about the very low metallicity of the selected galaxies.

In paper I (Kniazev et al. \cite{UM}) we reported a revision of
3 BCGs from the list of
Kunth \& \"Ostlin (\cite{KO}). For one, the low metallicity
value was confirmed while the other two showed an underestimation of 0.4~dex.
In this paper we present high S/N spectroscopy of two more BCGs (UM~408 and
A~1228+12) from the list by Kunth \& \"Ostlin (\cite{KO}), for which the
metallicity determination raised some concerns,  and  one
galaxy (UM~151) from the work by Telles (\cite{Telles96}). All these galaxies
were claimed as $Z \sim 1/20~Z_{\odot}$ objects based on the results of
earlier spectroscopy.
As a result of the previous and present work the number of extremely
metal-deficient BCGs from the Kunth \& \"Ostlin list has decreased by
$\sim$20\%.
The remaining objects are in fact representative of very metal-deficient
BCGs and deserve detailed multiwavelength studies.

\section{Observations and reduction}
\label{Obs}

\subsection{Observations}


\begin{table*}
\begin{center}
\caption{Journal of Observations}
\label{t:Tab1}
\begin{tabular}{lrrcccccc} \\ \hline \hline
\MC{1}{c}{ Object }     &
\MC{1}{c}{ Date }       &
\MC{1}{c}{ Exposure }   &
\MC{1}{c}{ Wavelength } &
\MC{1}{c}{ Dispersion } &
\MC{1}{c}{ Seeing }     &
\MC{1}{c}{ Airmass }     &
\MC{1}{c}{ PA }          \\

\MC{1}{c}{ }       &
\MC{1}{c}{ }       &
\MC{1}{c}{ time [s] }    &
\MC{1}{c}{ Range [\AA] } &
\MC{1}{c}{ [\AA/pixel] } &
\MC{1}{c}{ [arcsec] }    &
\MC{1}{c}{          }    &
\MC{1}{c}{ [degree] }     \\

\MC{1}{c}{ (1) } &
\MC{1}{c}{ (2) } &
\MC{1}{c}{ (3) } &
\MC{1}{c}{ (4) } &
\MC{1}{c}{ (5) } &
\MC{1}{c}{ (6) } &
\MC{1}{c}{ (7) } &
\MC{1}{c}{ (8) } & \\
\hline
\\[-0.3cm]
UM~151    & 1.02.2000   & 2x1800 & $ 3700\div7000$  & 0.81/0.54 &  1.8 &  1.50 &  0 \\
UM~408    & 2.02.2000   & 2x1800 & $ 3700\div7000$  & 0.81/0.54 &  1.4 &  1.30 &  6 \\
A~1228+12 & 20.01.2001  & 2x1800 & $ 3700\div7400$  & 2.4       &  1.7 &  1.18 &  74 \\
\hline \hline \\[-0.2cm]
\end{tabular}
\end{center}
\end{table*}

\begin{table*}[hbtp]
\centering{
\caption{Line intensities of the studied galaxies}
\label{t:Intens}
\begin{tabular}{lcccccc} \hline  \hline
\rule{0pt}{10pt}
& \MC{2}{c}{A1228+12} & \MC{2}{c}{UM151} & \MC{2}{c}{UM408}      \\ \hline
\rule{0pt}{10pt}
$\lambda_{0}$(\AA) Ion$^a$              & F($\lambda$)/F(H$\beta$)$^b$&I($\lambda$)/I(H$\beta$)$^c$ & F($\lambda$)/F(H$\beta$)$^b$&I($\lambda$)/I(H$\beta$)$^c$ & F($\lambda$)/F(H$\beta$)$^b$&I($\lambda$)/I(H$\beta$)$^c$ \\ \hline
3727\ [O\ {\sc ii}]\                      & 0.9503$\pm$0.0722 & 0.9259$\pm$0.0778   & 2.4257$\pm$0.1997 & 2.6919$\pm$0.2466  & 1.4502$\pm$0.1280 & 2.1700$\pm$0.2045 \\
3835\ H9\                                       & ---             & ---                   & ---             & ---            & 0.0590$\pm$0.0187 & 0.0860$\pm$0.0380 \\
3868\ [Ne\ {\sc iii}]\                    & 0.3791$\pm$0.0308 & 0.3693$\pm$0.0324       & ---             & ---              & 0.3812$\pm$0.0409 & 0.5390$\pm$0.0601 \\
3889\ He\ {\sc i}\ +\ H8\                 & 0.1634$\pm$0.0170 & 0.1958$\pm$0.0275   & 0.1262$\pm$0.0259 & 0.1815$\pm$0.0528  & 0.1474$\pm$0.0246 & 0.2082$\pm$0.0434 \\
3967\ [Ne\ {\sc iii}]\ +\ H7\             & 0.2254$\pm$0.0196 & 0.2555$\pm$0.0288   & 0.1020$\pm$0.0285 & 0.1470$\pm$0.0527  & 0.2077$\pm$0.0294 & 0.2842$\pm$0.0507 \\
4101\ H$\delta$\                          & 0.2210$\pm$0.0196 & 0.2492$\pm$0.0279   & 0.2409$\pm$0.0320 & 0.2947$\pm$0.0518  & 0.2113$\pm$0.0230 & 0.2749$\pm$0.0425 \\
4340\ H$\gamma$\                          & 0.4605$\pm$0.0360 & 0.4798$\pm$0.0416   & 0.4086$\pm$0.0389 & 0.4588$\pm$0.0553  & 0.4251$\pm$0.0362 & 0.5056$\pm$0.0492 \\
4363\ [O\ {\sc iii}]\                     & 0.1055$\pm$0.0130 & 0.1028$\pm$0.0131       & ---             & ---              & 0.0851$\pm$0.0170 & 0.1002$\pm$0.0202 \\
4471\ He\ {\sc i}\                        & 0.0321$\pm$0.0073 & 0.0313$\pm$0.0073   & 0.0271$\pm$0.0144 & 0.0273$\pm$0.0151  & 0.0266$\pm$0.0113 & 0.0302$\pm$0.0129 \\
4686\ He\ {\sc ii}\                       & 0.0403$\pm$0.0078 & 0.0392$\pm$0.0078       & ---             & ---                  & ---             & ---             \\
4713\ [Ar\ {\sc iv]}\ +\ He\ {\sc i}\     & 0.0208$\pm$0.0073 & 0.0203$\pm$0.0073       & ---             & ---                  & ---             & ---             \\
4713\ [Ar\ {\sc iv]}\                     & 0.0163$\pm$0.0080 & 0.0159$\pm$0.0080       & ---             & ---                  & ---             & ---             \\
4861\ H$\beta$\                           & 1.0000$\pm$0.0771 & 1.0000$\pm$0.0800   & 1.0000$\pm$0.0829 & 1.0000$\pm$0.0899  & 1.0000$\pm$0.0758 & 1.0000$\pm$0.0789 \\
4959\ [O\ {\sc iii}]\                     & 1.5651$\pm$0.1213 & 1.5251$\pm$0.1213   & 0.5509$\pm$0.0493 & 0.5270$\pm$0.0490  & 1.7714$\pm$0.1318 & 1.7174$\pm$0.1291 \\
5007\ [O\ {\sc iii}]\                     & 4.6938$\pm$0.3588 & 4.5736$\pm$0.3590   & 1.6881$\pm$0.1343 & 1.6069$\pm$0.1328  & 5.6848$\pm$0.4445 & 5.4322$\pm$0.4294 \\
5876\ He\ {\sc i}\                        & 0.0992$\pm$0.0107 & 0.0967$\pm$0.0109   & 0.1227$\pm$0.0211 & 0.1077$\pm$0.0194  & 0.1538$\pm$0.0311 & 0.1161$\pm$0.0238 \\
6300\ [O\ {\sc i}]\                       & 0.0230$\pm$0.0096 & 0.0224$\pm$0.0096   & 0.0636$\pm$0.0171 & 0.0540$\pm$0.0152  & 0.0584$\pm$0.0232 & 0.0399$\pm$0.0161 \\
6312\ [S\ {\sc iii}]\                     & 0.0146$\pm$0.0087 & 0.0143$\pm$0.0087   & 0.0161$\pm$0.0119 & 0.0137$\pm$0.0105  & 0.0309$\pm$0.0231 & 0.0211$\pm$0.0159 \\
6364\ [O\ {\sc i}]\                       & 0.0085$\pm$0.0055 & 0.0083$\pm$0.0055   & 0.0203$\pm$0.0145 & 0.0171$\pm$0.0128      & ---             & ---             \\
6548\ [N\ {\sc ii}]\                      & 0.0102$\pm$0.0068 & 0.0100$\pm$0.0068   & 0.1455$\pm$0.0234 & 0.1212$\pm$0.0206  & 0.0342$\pm$0.0306 & 0.0222$\pm$0.0201 \\
6563\ H$\alpha$\                          & 2.6763$\pm$0.2002 & 2.6217$\pm$0.2185   & 3.4333$\pm$0.2673 & 2.8819$\pm$0.2533  & 4.3736$\pm$0.3226 & 2.8264$\pm$0.2289 \\
6584\ [N\ {\sc ii}]\                      & 0.0310$\pm$0.0229 & 0.0302$\pm$0.0229   & 0.4404$\pm$0.0383 & 0.3661$\pm$0.0353  & 0.0957$\pm$0.0245 & 0.0615$\pm$0.0160 \\
6678\ He\ {\sc i}\                        & 0.0250$\pm$0.0071 & 0.0244$\pm$0.0071   & 0.0379$\pm$0.0144 & 0.0313$\pm$0.0124  & 0.0506$\pm$0.0220 & 0.0319$\pm$0.0140 \\
6717\ [S\ {\sc ii}]\                      & 0.0936$\pm$0.0109 & 0.0912$\pm$0.0113   & 0.4766$\pm$0.0409 & 0.3926$\pm$0.0378  & 0.2467$\pm$0.0342 & 0.1545$\pm$0.0222 \\
6731\ [S\ {\sc ii}]\                      & 0.0645$\pm$0.0096 & 0.0628$\pm$0.0099   & 0.3423$\pm$0.0328 & 0.2816$\pm$0.0298  & 0.1635$\pm$0.0260 & 0.1021$\pm$0.0168 \\
7065\ He\ {\sc i}\                        & 0.0173$\pm$0.0062 & 0.0168$\pm$0.0063       & ---             & ---                  & ---             & ---             \\
7136\ [Ar\ {\sc iii}]\                    & 0.0430$\pm$0.0083 & 0.0419$\pm$0.0085       & ---             & ---                  & ---             & ---             \\
  & & \\
C(H$\beta$)\ dex          & \MC {2}{c}{0.00$\pm$0.10} & \MC {2}{c}{0.20$\pm$0.10} & \MC {2}{c}{0.57$\pm$0.10} \\
EW(abs)\ \AA\             & \MC {2}{c}{2.45$\pm$1.08} & \MC {2}{c}{0.90$\pm$0.67} & \MC {2}{c}{0.05$\pm$0.94} \\
F(H$\beta$)$^d$\          & \MC {2}{c}{232$\pm$13}    & \MC {2}{c}{68$\pm$4}      & \MC {2}{c}{33$\pm$2}      \\
EW(H$\beta$,emis)\ \AA\   & \MC {2}{c}{93$\pm$5}      & \MC {2}{c}{20$\pm$2}      & \MC {2}{c}{50$\pm$3}      \\
\hline  \hline
\MC{7}{l}{~~} \\
\MC{7}{l}{$^a$ The rest-frame wavelength in \AA; $^b$ Observed flux ratio; $^c$ Corrected flux ratio; $^d$ in units of 10$^{-16}$ ergs\ s$^{-1}$cm$^{-2}$.}\\
\end{tabular}
 }
\end{table*}

The spectra of UM~151 and UM~408 were obtained with the TWIN spectrograph
attached to the Cassegrain focus of the 3.5\,m telescope at the
Calar Alto Observatory (Spain) as supplementary objects to the main program
devoted to the detailed spectroscopy of the HSS  (Hamburg/SAO Survey,
Ugryumov et al.~\cite{Ugryumov01}, and references therein) blue compact
galaxies.
Parameters of these observations are shown in Table
\ref{t:Tab1}. The setup used for  TWIN was
the T07 grating in second order for the blue and T06 in first
order for the red arm, that provided dispersions of 54~\AA~mm$^{-1}$ and
36~\AA~mm$^{-1}$ respectively. We have used the CCD detectors
SITE12a-11 and  SITe6a-11 for the blue and red arms with the 5500~\AA\ beam
splitter and a slit width of 2\farcs1 for UM~151 and 1\farcs2 for UM~408.
The resulting FWHM spectral resolution measured on strong lines
were 3.1~\AA\ and 2.5~\AA\  in the blue and red, for
UM~151, and 2.9 and 2.6~\AA, for UM~408. The scale along the
slit was 0\farcs56~pix$^{-1}$.


   \begin{figure*}
   \centering
   \includegraphics[angle=-90,width=11cm]{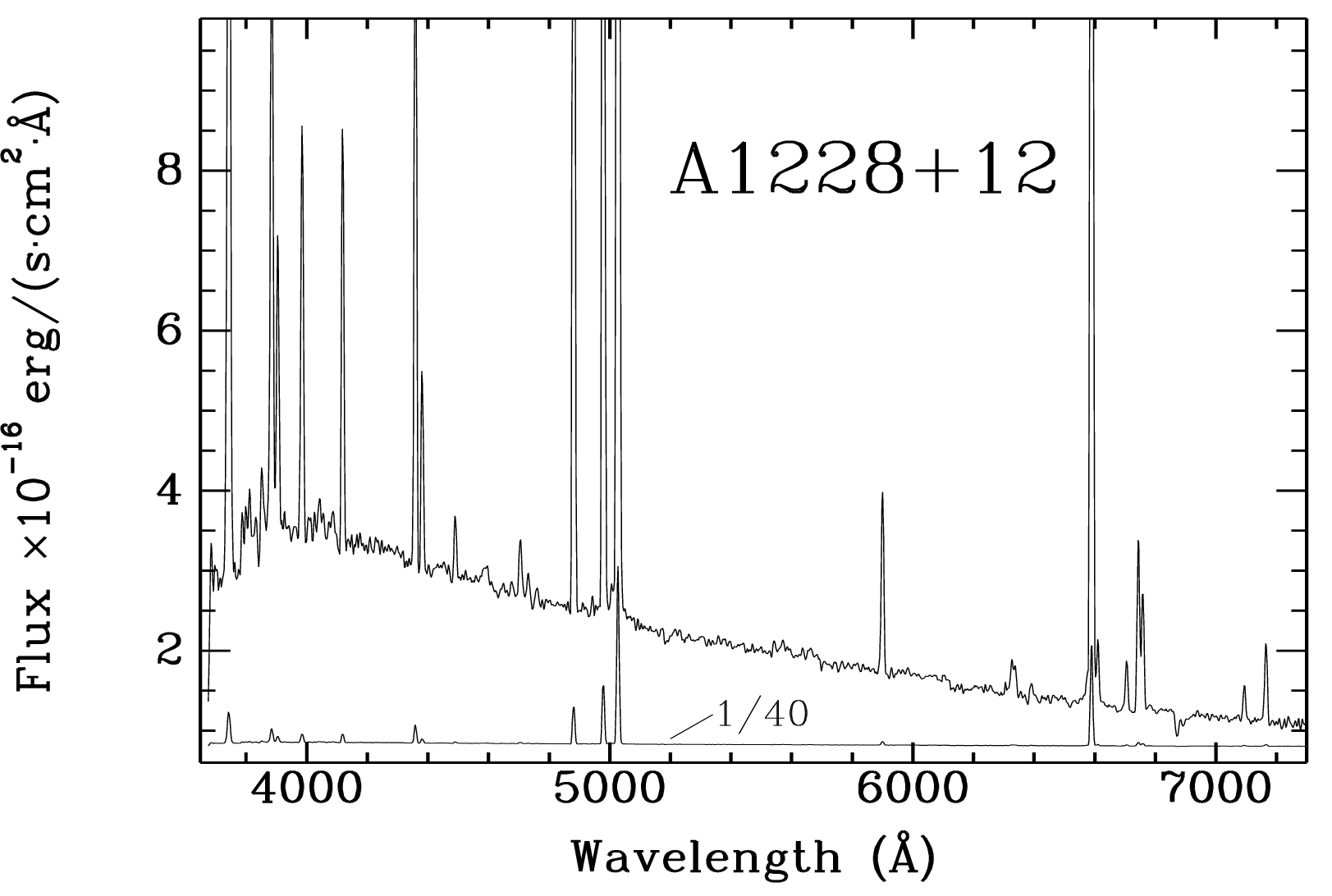}
   \includegraphics[angle=-90,width=11cm]{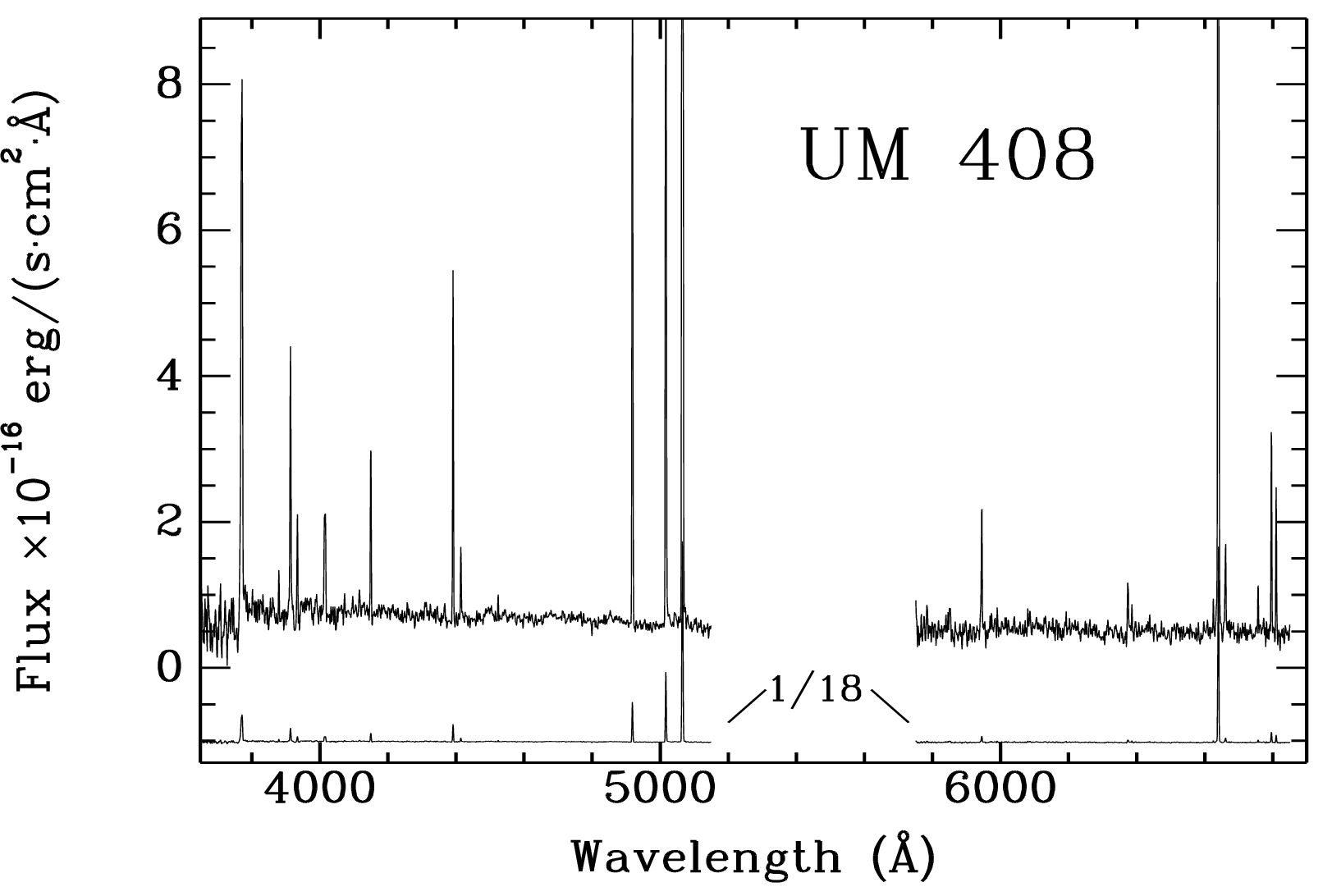}
   \includegraphics[angle=-90,width=11cm]{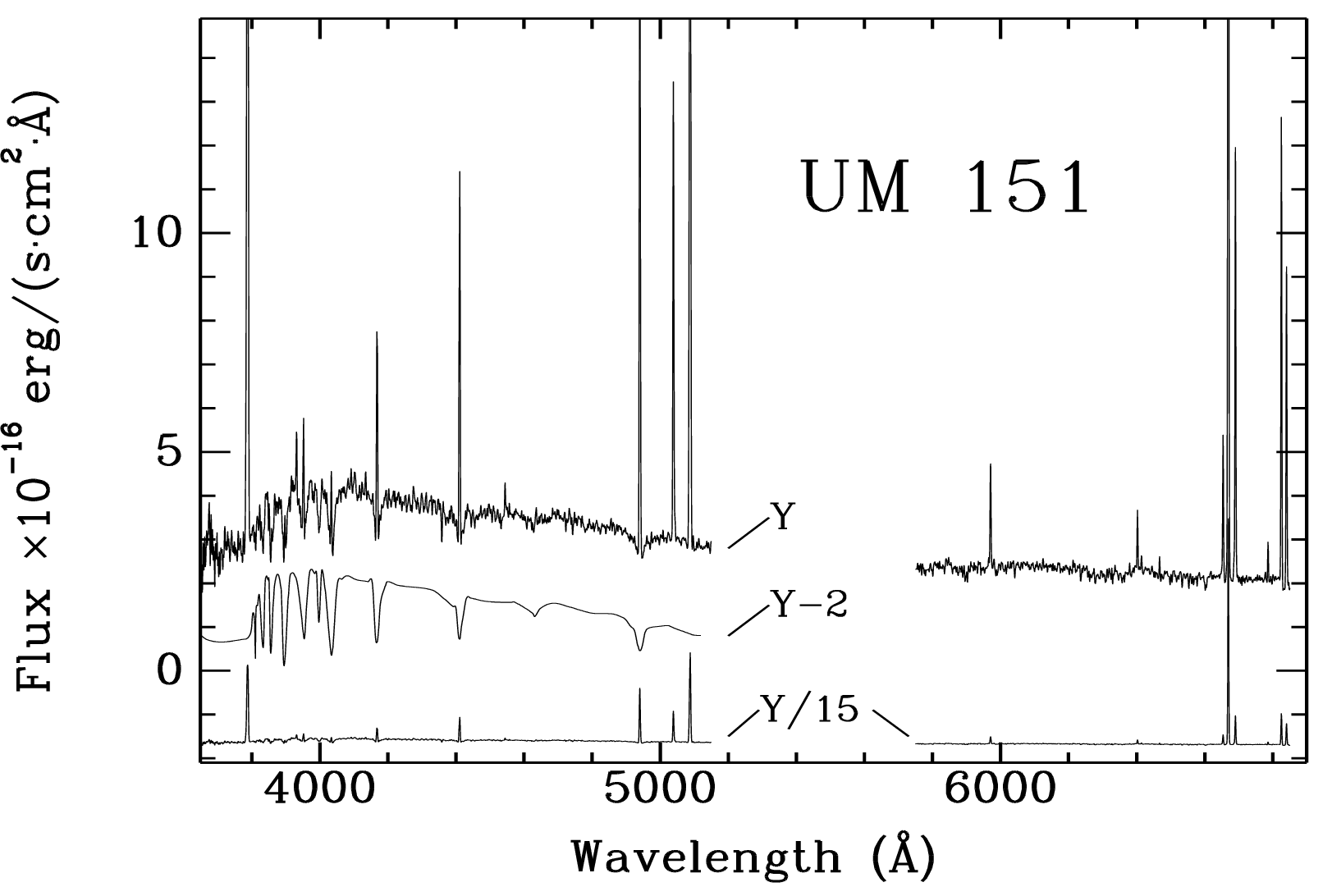}
    \caption{1D-spectra in the observed wavelength scale for the
     three galaxies discussed in the paper.
     In the bottom of each spectra the scaled down versions are drawn
     to show the relative intensities of strong lines. For UM~151
     the continuum with Balmer absorption lines is shown separately,
     shifted down along the ordinate (Y) axis by two flux units
     (marked as ``Y-2'').   }
	 \label{fig:Spectra}
   \end{figure*}

The spectroscopic data for A~1228+12 (RMB~132) were obtained  with the
6\,m telescope of the Special Astrophysical Observatory of the Russian
Academy of Sciences (SAO RAS).  Parameters of these observations
are shown in Table~\ref{t:Tab1}.
The long-slit spectrograph (LSS) (Afanasiev et al. \cite{Afanasiev95})
was used with a Photometrics CCD detector of 1024$\times$1024 pixels with a
$24\times24~\mu$m pixel size. Observations were
conducted mainly with the software package {\tt NICE} in MIDAS,
as described by Kniazev \& Shergin (\cite{Kniazev95}).
The scale along the slit was 0\farcs39~pix$^{-1}$.
A grating with 651~grooves~mm$^{-1}$ and a slit width of 2\arcsec\ were used,
giving a FWHM spectral resolution of 8~\AA. Two 0.5-hour spectra were
obtained, one after the other, each in the spectral range of
$3700-6100$~\AA\ and $5000-7400$~\AA, with the same
pointing and long slit orientation. Spectra were extracted from the same
region and the two spectra were combined to get the full spectrum of the
object for analysis.

For each night we obtained biases, flat-fields and illumination 
correction images. Comparison lamps of Fe-Ne and Ar-Ne-He were used for 
wavelength calibration for the Calar Alto and SAO data, respectively.
For flux calibrations, spectrophotometric standard stars from Bohlin
(\cite{Bohlin96}) for the 6\,m telescope observations and
Oke (1990) for the 3.5\,m telescope were used.
Average sensitivity curves were produced for each night with r.m.s.
deviations of $\sim$5\% in the whole spectral blue + red range.

\subsection{Reduction}

Standard reduction procedures were used
with the IRAF\footnote{IRAF: the Image Reduction and Analysis
Facility is
distributed by the National Optical Astronomy Observatories, which is
operated by the Association of Universities for Research in Astronomy,
In. (AURA) under cooperative agreement with the National Science
Foundation (NSF).}
package. Once 2D spectra were wavelength calibrated and sky 
subtracted, flux calibration was performed by using the average sensitivity
curves.

The 1D spectra were extracted with the apertures
of 3.6$\times$2.0\arcsec, 6.2$\times$2.1\arcsec\  and 6.2$\times$2.1\arcsec,
respectively for the galaxies A~1228+12, UM~151 and UM~408.
The 1D final spectra are shown in Fig.~\ref{fig:Spectra}.
The continuum determination and
the measurements of the flux and equivalent width (EW) of spectral lines
were performed with MIDAS\footnote{MIDAS is an acronym for the
European Southern Observatory package --- Munich Image Data Analysis System.}
(for details, see e.g., Kniazev et al. \cite{Kniazev00}).
EWs for individual emission lines were measured with the standard
MIDAS procedure {\tt INTEGRATE/LINE}.
The flux and equivalent width of the blended lines were measured
using  Gaussian decomposition fitting. In both cases the background was
drawn by two methods: manually and with the use of the automaticv procedure,
with the help of the algorithm, described in detail by Shergin et al.
(\cite{Shergin96}). The results of both cases were compared.
The errors of the sensitivity curve and those of the line
intensities have been combined in quadrature and propagated to calculate
element abundances.

In particular, for the A~1228+12 spectrum with $\sim$7~\AA\ resolution,
deblending was performed for H$\gamma$/[\ion{O}{iii}]$\lambda$4363,
H$\alpha$/[\ion{N}{ii}]$\lambda$6548,6584,
[\ion{O}{i}]$\lambda$6300/[\ion{S}{ii}]$\lambda$6312. With these procedures
the redshift and the line width were derived first for the stronger line of
the blend. For the gaussian fitting of the fainter blend components, these
parameters have been fixed with the values derived for the stronger component.
For the fitting of [\ion{N}{ii}]$\lambda$6548,6584 lines we
also fixed the intensity ratio of the two lines as 1:3, expected from
theory (e.g. Aller \cite{Aller84}). The uncertainties of these fitted values
were determined from
the residual noise of the spectrum near the lines under analysis. These
uncertainties were combined, as well as for all other measured line
intensities, with the other error components (see below).
Therefore, the derived errors of [\ion{N}{ii}] lines can be large, and
their intensity ratio in the table is theoretical. For the TWIN spectra the
spectral resolution is sufficient to measure each line separately.
While in the spectrum of UM~408 [\ion{N}{ii}]-lines are rather faint, their
line ratio is occasionally close to the theoretical value.
In addition to the noise of the underlying continuum,  quoted errors in the
line intensities include two more components:
one comes from Poisson statistics of photon flux, the other comes from the
uncertainties of the sensitivity curve, contributing a few per cent to all
lines. The line intensity errors presented in Table~\ref{t:Intens}
incorporate all three components, and thus should be reliable estimators
for other derived physical parameters and chemical abundances in the
\ion{H}{ii} regions considered. An independent check of the reliability of
the cited errors is the good consistency between our results and the results
of Kinman \& Davidson (\cite{KD}) for A~1228+12 (see section~\ref{RMB132}).
Both sets of line intensities are consistent within the cited errors,
if their extinction correction is accounted for. The latter can be
overestimated, since Kinman \& Davidson indicate a mismatch in the continuum
level for independent red and blue spectra. Another factor leading to
small differences is that Kinman \& Davidson did not account for the
underlying Balmer absorption. In the present work we determined underlying
Balmer absorption at H$\beta$ of $EW$(H$\beta$)$\sim$2.5\AA.
We derived from the spectrum of A~1228+12 the value of C(H$\beta$)=0. This is
consistent within the cited uncertainties ($\sigma_{C(H\beta)}$=0.10) with
the minimum value of C(H$\beta$)=0.043 following from the Galaxy extinction
in this direction, A$_\mathrm{B}=$0.12 mag (see Table~\ref{t:Param}).
We have checked the effect of the change of C(H$\beta$) from zero to 0.043
on the derived element abundances. The O/H value does not change at all.
The values of $log$ (N/O), (Ne/O), (S/O) and (Ar/O) change by only 0.02--0.03
dex, which is many times smaller than their cited uncertainties.

\begin{table}[tp]
\centering{
\caption{Abundances in the studied galaxies}
\label{t:Chem}
\begin{tabular}{lcc} \hline   \hline
\rule{0pt}{10pt}
Value                                & A~1228+12          & UM~408                  \\ \hline
$T_{\rm e}$(OIII)(K)\                & 16100$\pm$1100~~  & 14800$\pm$1400~~       \\
$T_{\rm e}$(OII)(K)\                 & 14300$\pm$900 ~~  & 13600$\pm$1200~~       \\
$T_{\rm e}$(SIII)(K)\                & 15100$\pm$900 ~~  & 14000$\pm$1200~~       \\
$N_{\rm e}$(SII)(cm$^{-3}$)\         &  10$\pm$10 ~~      &  10$\pm$10 ~~           \\
& & \\
O$^{+}$/H$^{+}$($\times$10$^{-5}$)\     & 0.936$\pm$0.186~~  & 2.518$\pm$0.689~~       \\
O$^{++}$/H$^{+}$($\times$10$^{-5}$)\    & 4.198$\pm$0.770~~  & 6.050$\pm$1.546~~       \\
O$^{+++}$/H$^{+}$($\times$10$^{-5}$)\   & 0.254$\pm$0.098~~  &   --                    \\
O/H($\times$10$^5$)\                 & 5.388$\pm$0.798~~  & 8.568$\pm$1.693~~       \\
12+log(O/H)\                         & ~7.73$\pm$0.06~~   & ~7.93$\pm$0.09~~        \\
& & \\
N$^{+}$/H$^{+}$($\times$10$^{-7}$)\     & 2.470$\pm$1.472~~  & 5.491$\pm$1.439~~       \\
ICF(N)\                              & 5.758              & 3.402                   \\
log(N/O)\                            & --1.58$\pm$0.27~~  & --1.66$\pm$0.14~~       \\
& & \\
Ne$^{++}$/H$^{+}$($\times$10$^{-5}$)\   & 0.737$\pm$0.143~~  & 1.369$\pm$0.381~~       \\
ICF(Ne)\                             & 1.284              & 1.416                   \\
log(Ne/O)\                           & --0.76$\pm$0.11~~  & --0.65$\pm$0.15~~       \\
& & \\
S$^{+}$/H$^{+}$($\times$10$^{-7}$)\     & 1.658$\pm$0.232~~  & 2.990$\pm$0.537~~       \\
S$^{++}$/H$^{+}$($\times$10$^{-7}$)\    & 7.212$\pm$4.560~~  & 13.490$\pm$10.660~~     \\
ICF(S)\                              & 1.709              & 1.357                   \\
log(S/O)\                            & --1.55$\pm$0.23~~  & --1.58$\pm$0.29~~       \\
& & \\
Ar$^{++}$/H$^{+}$($\times$10$^{-7}$)\   & 1.500$\pm$0.327~~  & ---   \\
Ar$^{+++}$/H$^{+}$($\times$10$^{-7}$)\  & 1.957$\pm$1.023~~  & ---   \\
ICF(Ar)\                             & 1.026              & ---   \\
log(Ar/O)\                           & --2.18$\pm$0.15~~  & ---   \\
\hline  \hline
\end{tabular}
 }
\end{table}


\begin{table}
\caption{\label{Tab1} Main parameters of studied galaxies}
\label{t:Param}
\begin{tabular}{lccc} \\ \hline \hline
Parameter                      & UM~151                 & UM~408                    & A~1228+12     \\ \hline
$\alpha_\mathrm{2000}$         & ~~01 57 38.87          & ~~02 11 23.55             & ~~12 30 48.52 \\
$\delta_\mathrm{2000}$         & +02 25 23.9            & +02 20 31.0               & +12 02 42.1   \\
A$_\mathrm{B}$$^N$             & 0.12                   & 0.15                      & 0.12          \\
B$_\mathrm{tot}$               & 16.21$^{(1)}$          & 17.74$^{(1)}$             & 17.15$^{(3)}$   \\
V$_\mathrm{Hel}$ (\kms)        & 4851$^{(2)}$           & 3507$^{(4)}$              & 1263$^{(5)}$    \\
Dist (Mpc)                     & 64.7                   & 46.8                      & 17.0$^{V}$    \\
M$_\mathrm{B}^0$\,$^{(6)}$     & --17.96                & --15.76                   & --14.10       \\
Opt. size (\arcsec)$^{(7)}$    & 35$\times$15.3$^{(1)}$ & 15.6$\times$10.6$^{(4)}$  & 12$\times$9$^{(3)}$    \\
Opt. size (kpc)                & 11.0$\times$2.4$^{(2)}$ & 3.5$\times$2.4$^{(2)}$   & 1.0$\times$0.75$^{(2)}$ \\
12+log(O/H)  \                 & 8.5$^{(2)}$~           & 7.93$^{(2)}$              & 7.73$^{(2)}$    \\
\ion{H}{i} flux$^{(8)}$        &  $<$1.2$^{(9)}$        & 1.5$^{(4)}$               & 1.4$^{(5)}$     \\
W$_\mathrm{50}$ (\kms)         &  ---                   & 77$^{(4)}$                & 84$^{(5)}$      \\
M(\ion{H}{i}) (10$^{(8)} M_{\odot}$)  & $<$11.9$^{(2)}$ & 7.8$^{(2,4)}$             & 0.95$^{(2,5)}$   \\
M(\ion{H}{i})/L$_\mathrm{B}$$^{(10)}$  & $<$2.0$^{(2)}$ & 2.5$^{(2)}$               & 1.5$^{(2)}$     \\ \hline\hline

\multicolumn{4}{l}{(1)  -- Salzer et al. (\cite{Salzer89b})($V_{25}$-isophote, $b/a$ - } \\
\multicolumn{4}{l}{~~~~~~~ minor-to-major axis ratio, from LEDA)} \\
\multicolumn{4}{l}{(2)  -- parameters derived in this paper.} \\
\multicolumn{4}{l}{(3)  -- Binggeli \& Cameron~(\cite{BC93})} \\
\multicolumn{4}{l}{(4)  -- Smoker et al.~\cite{SDAM00}} \\
\multicolumn{4}{l}{(5)  -- Staveley-Smith et al.~\cite{Staveley92}} \\
\multicolumn{4}{l}{(6)  -- corrected for the Galaxy extinction.} \\
\multicolumn{4}{l}{(7)  -- $a \times b$ at  $\mu_\mathrm{B} =$25 mag/sq.arcsec. See note (1).} \\
\multicolumn{4}{l}{(8)  -- in units of (Jy$\cdot$\kms).} \\
\multicolumn{4}{l}{(9)  -- Thuan et al.~\cite{TLMP99}; upper limit is estimated} \\
\multicolumn{4}{l}{~~~~~~~~for W$_\mathrm{50}=100$~\kms} \\
\multicolumn{4}{l}{(10)  -- in units of ($M/L_\mathrm{B}$)$_{\odot}$.} \\
\multicolumn{4}{l}{(V)  -- accepted for the Virgo cluster (Tikhonov et al.~\cite{Tikhonov00})} \\
\multicolumn{4}{l}{(N)  -- data from NED, Schlegel et al. \cite{Schlegel98}.} \\
\end{tabular}
\end{table}

\section{Results of the abundance determination}
\label{Results}

Relative intensities of all emission lines together with the equivalent width
$EW$(H$\beta$,emis), extinction coefficient $C$(H$\beta$) and
the equivalent width of the hydrogen absorption lines
are given in Table~\ref{t:Intens}.
The extinction coefficient $C$(H$\beta$) was derived from the
hydrogen Balmer emission decrement using the self-consistent method described
by Izotov et al.~(\cite{Izotov94}). For both UM~408 and A~1228+12 the
continuum was drawn as a running mean without accounting for possible
absorption features.
For UM~151, before obtaining emission line intensities, the underlying
continuum was drawn including Balmer absorption lines and other apparent
absorption features. Their measured equivalent widths were used for age
estimates (see Table \ref{Bal_abs}).
The quoted $EW(abs)$ for this galaxy in Table~\ref{t:Intens} is the
residual value that is derived after the underlying continuum was drawn,
including strong Balmer absorptions. The $EW$ of emission H$\beta$ is
calculated on a running mean continuum that will be compared with the model
value below.
The derived extinction coefficients are in the range from
zero for A~1228+12, 0.2 for UM~151, to $\sim$0.6 for UM~408.
The latter value
is somewhat larger than usually is observed in this type of galaxies.
For A~1228+12 and UM~408 the chemical abundances and physical
parameters have been obtained with the method outlined in the paper of
Izotov et al.~(\cite{Izotov97}). The resulting values are given in
Table \ref{t:Chem}.

For UM~151, since no measurable [\ion{O}{iii}] $\lambda$4363~\AA\ line
has been detected, the metallicity has been estimated by means of empirical
methods (see more details in section~\ref{UM151}).

The $EW$ of emission H$\beta$ presented in Table~\ref{t:Intens} were
used to derive starburst ages according to the Starburst99 model (Leitherer
et al. \cite{Starburst99}) in sections \ref{UM151} and \ref{RMB132}. It was
assumed that extinction values for the ionized gas and the young
stellar clusters are similar.

\section{Discussion and summary}
\label{Discussion}

We discuss below the individual galaxies in more detail, based mainly
on the new spectral data, and the observational data from the
literature, appropriate for the present discussion. The main
parameters of the studied galaxies are given in Table \ref{t:Param}.

\subsection{UM~151=Mkn~1169}
\label{UM151}

This galaxy does not look like a bona fide BCG. Its appearance resembles
that of a face-on disk with somewhat disturbed outermost parts, but
clearly without spiral arms. Salzer (\cite{Salzer89a}) classified it
as a Dwarf Amorphous Nuclear Starburst (DANS).
Its absolute magnitude quoted here (Table~\ref{t:Param},
$M_B=-$17\fm96) is  within the range for DANS.
A bright knot is seen near the galaxy center, but
does not change the generally regular appearance of this galaxy.
The low metallicity value from Telles (\cite{Telles96}) was derived from a
very low S/N spectrum (see Terlevich et al. \cite{Terlevich91}).

For the metallicity estimation both the Pilyugin (\cite{Pil01}) calibration 
based on the strong oxygen lines and the recently reported calibration based
on the [\ion{N}{ii}] line (Denicol\'o et al.~\cite{DTT02})
give very close values of $12+log(O/H)$, 8.50 and 8.47, respectively.

Balmer absorption lines and other features with reliable detections
can be used to estimate the age of star-formation episodes.

The Equivalent Widths ($EW$) of Balmer absorption lines from the underlying
continuum have been calculated following the prescription in
Gonz\'alez-Delgado et al. (\cite{Rosa}) and the results are presented in
Table~\ref{Bal_abs}. The comparison with the Gonz\'alez-Delgado et al.
(\cite{Rosa}) models gives an age for the starburst consistent with an
instantaneous starburst $\sim$10 Myr old. The $EW$ of H$\beta$ emission
is also consistent with the age of instantaneous starburst of $\sim$10 Myr
(Leitherer et al. \cite{Starburst99}).
Following Raimann et al. (\cite{Raimann00}), we also measured
the $EW$ of the \ion{Ca}{ii} K-line and the G-band as well as
the continuum flux ratios in the 3 bands (see Table~\ref{Bal_abs}).
All but one of the parameters are consistent with a mixture of two
starbursts with ages $\sim$10~Myr and
a $\sim$50~Myr. The rather high $EW$ of the G-band suggests
an additional
contribution of a stellar population with an age of a few hundred Myr.

The known upper limit on \ion{H}{i} flux of this galaxy (see Table
\ref{t:Param}) is rather high, so the upper limit on the ratio $M(HI)/L_B$
is consistent with the range typical of gas-rich starbursting galaxies.


\begin{table}
\begin{center}
\caption{The absorption lines $EW$ and continuum ratios of UM~151.}
\label{Bal_abs}
\begin{tabular}{lr|lr} \\ \hline \hline
\MC{1}{c}{ Abs. line}     &
\MC{1}{c|}{ Value }        &
\MC{1}{c}{ Band }        &
\MC{1}{c}{ Value }        \\
\hline
\\[-0.3cm]
EW(H$_{\beta}$) & 5.4   & CaII-K & 2.2 \\
EW(H$_{\gamma}$) & 4.0  & G band & 3.2 \\
EW(H$_{\delta}$) &  6.1 & F$_{3660}$/F$_{4020}$ & 0.71 \\
EW(H$_{8}$) & 7.4  	& F$_{3780}$/F$_{4020}$ & 0.88 \\
EW(H$_{9}$) & 7.9       & F$_{4510}$/F$_{4020}$ & 0.83 \\
EW(H$_{10}$) & 4.0 & &\\
\hline \hline \\[-0.2cm]
\end{tabular}
\end{center}
\end{table}

\subsection{UM~408}

Appearing like a typical blue compact dwarf, this galaxy was classified by
Salzer (\cite{Salzer89a})  as a Dwarf \ion{H}{ii} Hotspot Galaxy. The absolute
$B$-band luminosity is $M_{B}=-15.16$ (Campos-Aguilar et al. \cite{Campos}),
almost
at the lower end of the BCG luminosity distribution, with a quoted diameter
of 2.1 kpc. The calculated metallicity reported by Masegosa et al.
(\cite{Masegosa94}) was 12+$log$(O/H) = 7.63 using the data from the
Spectroscopic
Catalogue of \ion{H}{ii} Galaxies (Terlevich et al.~\cite{Terlevich91}).
Using the same set of data Telles (\cite{Telles96}) reported a value of
12+$log$(O/H) = 7.66. The new estimation
with the present data suggests a significantly higher metallicity  with a
difference of 0.2 to 0.3~dex. This difference cannot be attributed to the
reddening estimation or differences in slit positioning. Comparing the
present data with Masegosa et al. (\cite{Masegosa94}) and Telles
(\cite{Telles96}),  the reddening coefficients are similar within the
uncertainties and the values of both $EW$(H$\beta$,emis) and integrated
H$\beta$ flux are the same. Therefore the main
reason for the discrepancy must be a poor estimation of the
[\ion{O}{iii}] $\lambda$4363 due to low S/N.
The comparison between the line intensities of both sets of data shows
that a large difference is found not only in the faint [\ion{O}{iii}] line,
but also in [\ion{O}{ii}] being larger by a factor of 2 for the TWIN data.
The same is also true for the measured [\ion{N}{ii}] line. The low S/N on
the continuum for this faint galaxy can account for the large difference
in oxygen abundance.

The measured \ion{H}{i} flux of this BCG (see Table \ref{t:Param}) corresponds
to a very high value of the parameter $M(HI)/L_B$. The $M(HI)/L_B$
is comparable to values derived for the most extreme objects in the sample of
dwarf galaxies with extended \ion{H}{i} (van Zee et al.~\cite{Zee95}).

\subsection{A~1228+12 = RMB~132 = VCC~1313}
\label{RMB132}

This galaxy is one of the most compact, almost starlike in appearance,
of the BCG family (Drinkwater \& Hardy~\cite{DH91}).  It resides
in the Virgo cluster, and because of the surrounding environment, its
properties are probably somewhat affected by more frequent interactions with
surrounding galaxies and the hot intracluster medium (ICM).
There is a number of sufficiently massive candidate  galaxies in the vicinity
of this BCG, which could trigger its current SF burst, including NGC~4478 at
the projected distance of 18.7\arcmin\ ($\sim$90 kpc) and M~87 at 20.8\arcmin\
($\sim$100 kpc), whose relative radial velocities are lower than 100~\kms.

Of the three galaxies studied in this paper, this is the faintest
system with an absolute $B$ magnitude of $-14.1$. Taking into account its
compactness and luminosity, this is the type of galaxy classified
as a Searle-Sargent object by Salzer (\cite{Salzer89a}), or \ion{H}{ii}
galaxies by Campos-Aguilar et al. (\cite{Campos}). This galaxy was
one of the first BCGs studied for metallicity purposes (Kinman \& Davidson
\cite{KD}, hereafter KD81). From
this study the metallicity reported of $12+log(O/H)=7.64\pm0.07$ is in
reasonable agreement with the present value of $7.73\pm0.06$, based on the
new, higher S/N data from the 6\,m telescope. The agreement can be
considered as a good one, taking into account that
the areas sampled are 29 arcsec$^2$ for KD81 and 7.2 arcsec$^2$ in this work.

However, it is evident that the abundance calculations by KD81 could have
some systematic differences, since
these authors did not account for temperature gradients. They used one
temperature for all zones, and noticed that this would lead to slight
underestimation of oxygen and neon abundances. Therefore we recalculated
oxygen and neon abundances from their relative intensities and the cited
errors using our methodology. This resulted in an upward shift of 0.04
dex for the KD81 oxygen abundance, giving
$12+log(O/H)=7.68\pm0.10$ and $log(Ne/O)=-$0.75. The difference between the
KD81 O/H abundance and the recalculated one appears to be
from a somewhat larger abundance of $O^{+}$ ions: $12+log(O^{+}/H)$ = 6.87
from KD81, and 7.05 in the model with the lower $T(O^{+})$.

One of the interesting features in the spectrum of A~1228+12 is the
appearance of an emission line
centered at $\lambda$4591~\AA, with flux $\sim$(2.5$\pm$1.2)\% of
H$\beta$ and FWHM=28~\AA. The only reasonable identification is as the
\ion{Si}{iii}~$\lambda$4565 line, characteristic of WR stars. The implied
radial velocity of this feature is then $\sim$450~\kms\ higher than the
system velocity of the galaxy found in other (narrow) emission lines.
Given the large line width and low S/N ratio, the velocity shift
is likely not significant. However the appearance of WR stars in the
starburst region of this BCG would not be unexpected.
Indeed, the observed $EW$ of H$\beta$ (93~\AA) according to Fig.85e
(corresponding to metallicity $z=$0.001, the closest to the oxygen abundance
of A~1228+12) from Starburst99 (Leitherer et al. \cite{Starburst99}),
corresponds to an instantaneous starburst of age 3--3.5 Myr. We adopted the
Salpeter IMF with $M_\mathrm{low}=0.1~M$\sunn\ for this estimate.
For the metallicity of $z=0.001$ this is exactly the age range where the
models predict significant numbers of WR-stars (0.5 to 2.5~\% of the number of
O-stars, see Schaerer \& Vacca \cite{SV98}).
Since we did not detect other characteristic WR features
(\ion{N}{iii}/\ion{N}{iv}~$\lambda$4640 and broad component of
\ion{He}{ii}~$\lambda$4686), that should have comparable $EW$, doubt about
the reality of the \ion{Si}{iii}~$\lambda$4565 line remains. Deeper
spectroscopy is necessary in order to measure the strength of the probable
WR features in this BCG.

The $M(HI)/L_B$ ratio for this BCG is quite large (see Table
\ref{t:Param}). Accounting for very large $EW$ of strong emission lines and
respective significant brightening ($\Delta B \gtrsim ~$1\fm0), this implies
a large gas mass-fraction. The latter is difficult to understand for a
low-mass
galaxy affected both by the ICM ram pressure (unless the BCG is only now
entering the hot gas) and tidal interactions from massive neighbours.

In fact, the very existence of such a low-metallicity BCG in a dense
environment such as the Virgo Cluster (despite the opposite tendency of
additional enrichment by heavy elements for the Virgo cluster BCGs, noticed by
Izotov \& Guseva \cite{IG89}), poses interesting questions on the
evolutionary history of this and similar objects. It is worth mentioning that
there are two more BCGs with well determined low metallicities
($Z\sim Z_{\odot}/20$) probably belonging to the Virgo Cluster or its
outskirts:
the Optical Counterpart of HI~1225+01 (Salzer et al. \cite{Salzer91}) and
VV~432=IC~3105=VCC~241 (Zasov et al. \cite{Zasov00}). Less intriguing, but
also not well understood, is the appearance of SBS~0335$-$052 (E+W), the pair
of extremely metal-deficient BCGs, situated at the outskirts of the loose
galaxy
group LGG 103 (Pustilnik et al. \cite{PBTLI}, Peebles \cite{Peebles01}).

\subsection{Conclusions}

We can summarize the results as follows:

\begin{itemize}
\item 
{New high S/N spectra of galaxies UM~151, UM~408 and A~1228+12 have been 
presented and their metallicities revisited. For A~1228+12  the new
value is a bit larger, but consistent within the uncertainties with
previous data. For the other two galaxies, significantly
higher metallicities have been estimated (5$\sigma$ and 3$\sigma$,
respectively, upward). None of the three galaxies
can be included in the sample of extremely low-metallicity BCGs ($Z
\lesssim 1/20~Z$\sunn).}
\item
{More precise abundances of nitrogen, neon and sulfur are derived for
  UM~408 and A~1228+12.}
\end{itemize}

Results from this paper and the previous work by Kniazev et al.
(\cite{UM}) showed an upward revision of metallicity in some of the
most metal-poor galaxies known. High S/N data of a statistically significant
galaxy sample are a prerequisite for studying the group properties of these
intriguing objects.

\begin{acknowledgements}

We acknowledge the partial support from INTAS grant 96-0500 and
Russian state program "Astronomy". The authors thank Y.Izotov for his help
in TWIN data reduction and H.Lee for useful comments, suggestions and
English corrections.
The authors are grateful to the referee
C.Leitherer for constructive criticism and useful suggestions, which helped
to improve the paper.
S.A.P. acknowledges the financial support from the Junta de
Andaluc\'{\i}a for a visit to the Instituto de Astrof\'{\i}sica de
Andaluc\'{\i}a, where part of the work on the paper was performed.
J.M. and I.M. acknowledge financial support by the
Spanish DGICYT, under the programs PB98-0521 and AYA2001-2089. This
research has made use of the NASA/IPAC Extragalactic Database (NED)
which is operated by the Jet Propulsion Laboratory, California
Institute of Technology, under contract with the National Aeronautics
and Space Administration. The use of the Digitized Sky Survey (DSS-II)
and the APM Database (IoA, Cambridge) is gratefully acknowledged.

\end{acknowledgements}

\end{document}